\documentclass[letterpaper,10pt]{article} 
\makeatletter
\usepackage{cite}
\usepackage{setspace}
\usepackage{multirow}
\usepackage{setspace}
\usepackage{array}
\usepackage{caption2}
\usepackage{graphicx}
\usepackage{booktabs}
\usepackage[namelimits]{amsmath} 
\usepackage{amssymb}             
\usepackage{amsfonts}            
\usepackage{mathrsfs}
\usepackage{makecell}
\usepackage{soul}
\usepackage{url}
\usepackage{arydshln}
\usepackage{booktabs}
\graphicspath{{figures/}}
\usepackage[left=2cm,right=2cm,top=2.5cm,bottom=2.5cm]{geometry}
\usepackage{osameet3} 

\newcommand\authormark[1]{\textsuperscript{#1}}

\usepackage{amsmath,amssymb}
\usepackage[colorlinks=true,bookmarks=false,citecolor=blue,urlcolor=blue]{hyperref} 

\begin{document}

\title{Brachial Plexus Nerve Trunk Segmentation Using Deep Learning: A Comparative Study with Doctors' Manual Segmentation}


\author{Yu Wang\authormark{1}, Binbin Zhu\authormark{2}, Lingsi Kong\authormark{3}, Jianlin Wang\authormark{3}, Bin Gao\authormark{2}, Jianhua Wang\authormark{2}, Dingcheng Tian\authormark{1}, and Yudong Yao\authormark{1,*}}

\address{\authormark{1}Research Institute for Medical and Biological Engineering, Ningbo University, China\\
\authormark{2}Department of Anesthesiology, Affiliated Hospital of Medical School of Ningbo University, Ningbo University, China\\
\authormark{3}Department of Anesthesiology, Ningbo No. 6 Hospital, Ningbo, China}

\email{\authormark{*}yaoyudong@nbu.edu.cn} 

\maketitle
		\textbf{Abstract: } Ultrasound-guided nerve block anesthesia (UGNB) is a high-tech visual nerve block anesthesia method that can observe the target nerve and its surrounding structures, the puncture needle's advancement, and local anesthetics spread in real-time. The key in UGNB is nerve identification. With the help of deep learning methods, the automatic identification or segmentation of nerves can be realized, assisting doctors in completing nerve block anesthesia accurately and efficiently. Here, we establish a public dataset containing 320 ultrasound images of brachial plexus (BP). Three experienced doctors jointly produce the BP segmentation ground truth and label brachial plexus trunks. We design a brachial plexus segmentation system (BPSegSys) based on deep learning. BPSegSys achieves experienced-doctor-level nerve identification performance in various experiments. We evaluate BPSegSys' performance in terms of intersection-over-union (IoU), a commonly used performance measure for segmentation experiments. Considering three dataset groups in our established public dataset, the IoU of BPSegSys are 0.5238, 0.4715, and 0.5029, respectively, which exceed the IoU 0.5205, 0.4704, and 0.4979 of experienced doctors. In addition, we show that BPSegSys can help doctors identify brachial plexus trunks more accurately, with IoU improvement up to 27\%, which has significant clinical application value.

\section{Main}
	Peripheral nerve block anesthesia is a type of local anesthesia by injecting local anesthetics to block the nerve conduction function in a specific body part of a patient. The critical step of peripheral nerve block anesthesia is nerve identification. In the past, blind guidance of needle and nerve stimulation are often used~\cite{10.1097/ALN.0b013e31828f4d80}. Such methods have strict requirements for doctors' experience and require real-time patient feedback, which are difficult to achieve and prone to complications
~\cite{tsui2010ultrasound}. With the development of ultrasound technologies, ultrasound-guided nerve block anesthesia (UGNB) techniques have begun to emerge in recent years~\cite{marhofer2007ultrasound}. With ultrasound equipments, doctors can improve the ability to identify nerve structures, observe the needle route and the spread of anesthetic drugs in real-time, reduce the difficulty of nerve block anesthesia, and improve the quality of anesthesia. However, UGNB require significant skills, and a lot experiences must be accumulated through practice to master the technique gradually
~\cite{marhofer2010current}. In practice, incorrect nerve identification may cause nerve injury.
	
	In order to assist doctors in identifying nerves and completing nerve block operations accurately and quickly, many researchers have investigated automatic nerve identification methods. Researchers have used traditional machine learning (ML) methods to identify nerves~\cite{hadjerci2014nerve}\cite{hadjerci2015nerve}\cite{hadjerci2016computer}. The traditional ML methods mainly include four technical steps: (1) ultrasonic image preprocessing, (2) feature extraction, (3) feature selection, and (4) feature classification. The primary purpose of ultrasound image preprocessing is to reduce the noise of ultrasound images and enhance the contrast between different tissue structures. Feature extraction mainly uses manual methods to extract the texture features, morphological features, and gray value features of the images. Subsequently, the extracted features are sent to a machine learning model for feature selection and feature classification to accomplish nerve identification. However, in feature processing in traditional ML methods, manual approaches are often used, which only extract and select image features in the lower or middle dimensions and it can not achieve satisfactory results.
	
	In recent years, deep learning has thrived in the medical ultrasound imaging field and has achieved more and more impressive results, such as tumor benign and malignant classification~\cite{shi2016stacked}, carotid artery segmentation~\cite{zhou2020voxel}, and breast lesion detection~\cite {yap2017automated}. Compared with traditional ML methods, deep learning automatically extracts and selects high-dimensional features of images, and analyzes the features to complete specific tasks. At present, there are many nerve identification research activation based on deep learning, mainly using nerve segmentation to identify nerves~\cite{zhao2017improved}\cite{ding2020multiple}\cite{van2021hybrid}\cite{wang2019ultrasound}\cite{wang2021segmentation}\cite{zhang2017image}\cite{liu2018segmentation}\cite{huang2019applying}. Zhao et al.~\cite{zhao2017improved} used modified U-Net as a deep learning network and used 5635 ultrasound images to study the segmentation of the brachial plexus region. The modified U-Net has a dice coefficient that is only 0.005 lower than the original U-Net~\cite{ronneberger2015u} when the quantity of parameters are much fewer than the original U-Net. Ding et al.~\cite{ding2020multiple} used Brachial Plexus Multi-instance Segmentation Network (BPMSegNet) to segment the nerves, arteries, veins, and muscles in the brachial plexus ultrasound images. They used average precision (AP) as the performance metrics and achieved an AP of 37.62. Other studies include~\cite{van2021hybrid}\cite{wang2019ultrasound}\cite{wang2021segmentation}\cite{zhang2017image}\cite{liu2018segmentation}\cite{huang2019applying}. In Table \ref{other papers}, we summarize the studies on nerve segmentation using deep learning methods in recent years. Notice that, we find two limitations in those studies. First, all studies segmented a \textbf{\emph{nerve block region}} rather than individual \textbf{\emph{nerve trunks}}. The segmentation of a nerve block region can help doctors to roughly locate the appropriate nerve position, but the doctors still need to search for the nerve based on the result of the segmentation region. In contrast, individual nerve trunks segmentation can directly help doctors locate nerve trunks and improve nerve block anesthesia's accuracy and efficiency. Second, previous nerve identification research has focused on the comparison between different artificial intelligence (AI) algorithms, without investigating whether the algorithms could actually assist doctors in improving nerve identification performance.

\begin{table}[h]
\begin{center}
\begin{minipage}{\textwidth}
		\caption{Summary of nerve segmentation using deep learning methods.}
		\resizebox{\textwidth}{!}{
		\begin{tabular}{lllll}
			\toprule
			Nervous Category                 & Dataset                                & Data Annotation Method                   & Network Structure               & Paper        \\ \midrule
			Brachial plexus & Kaggle public dataset &Brachial plexus \emph{region} & Modified U-Net & \cite{zhao2017improved}           \\
			&                                        &                                          &                                 & \cite{van2021hybrid}            \\
			&                                        &                                          &                                 & \cite{wang2019ultrasound}           \\
			&                                        &                                          &                                 & \cite{wang2021segmentation}             \\
			& Private dataset       & Brachial plexus \emph{region} & Modified U-Net                  & \cite{zhang2017image}            \\
			&                                        &                                          & BPMSEGNET                       & \cite{ding2020multiple}            \\
			&                                        &                                          & FCN                             & \cite{liu2018segmentation}            \\
			Femoral nerve                    & Private dataset                        & Femoral nerve \emph{region}                  & U-Net                           & \cite{huang2019applying}            \\ \hdashline
			\textbf{Brachial plexus}                 & \textbf{Private dataset}                         & \textbf{Brachial plexus \emph{trunks}}      & \textbf{Attention U-Net }               & \textbf{Our research} \\ \bottomrule
		\end{tabular}}
		\label{other papers}
\end{minipage}
\end{center}
\end{table}

	In this study, we aim to address the above two problems and advance the clinical application of AI algorithms in UGNB. To this end, our research includes the following tasks and contributions. (1) While the datasets used in previous work are all segmentation of nerve block regions, we create a new brachial plexus trunk segmentation dataset (BPSegData). We collect 320 brachial plexus ultrasound images (BPUS) from the Affiliated Hospital of Ningbo University and Ningbo No. 6 Hospital and annotate the brachial plexus trunk on BPUS. First, three experienced anesthesiologists independently label each brachial plexus trunk on BPUS, which can be used to measure each doctor's identification accuracy of the brachial plexus trunk. Then, three anesthesiologists work together as a team to label the brachial plexus trunk, which is used as the ground truth for the brachial plexus trunk (i.e., BPSegData). (2) Using our BPSegData, we develop a deep learning system for automatic segmentation of brachial plexus nerve trunks  (i.e., BPSegSys) and compare the deep learning segmentation results with each anesthesiologist's work to examine the clinical applicability of BPSegSys. (3) To investigate whether BPSegSys can be used to improve clinical work, one anesthesiologist labels the brachial plexus trunk on the BPUS again based on the BPSegSys results. Afterwards, the anesthesiologist's two labeling results are compared (the anesthesiologist's initial labeling and the second labeling based on BPSegSys). Overall, our research results show that BPSegSys reaches the level of brachial plexus trunk identification performance comparable to that of experienced doctors. In addition, BPSegSys is shown to be able to assist doctors in segmentating brachial plexus trunk with improved accuracy, which demonstrates the significant potential in clinical applications.
	\section{Results}
	\subsection{Image and system overview}
\begin{figure*}[h]
		\centering
		\includegraphics[width=11.5cm]{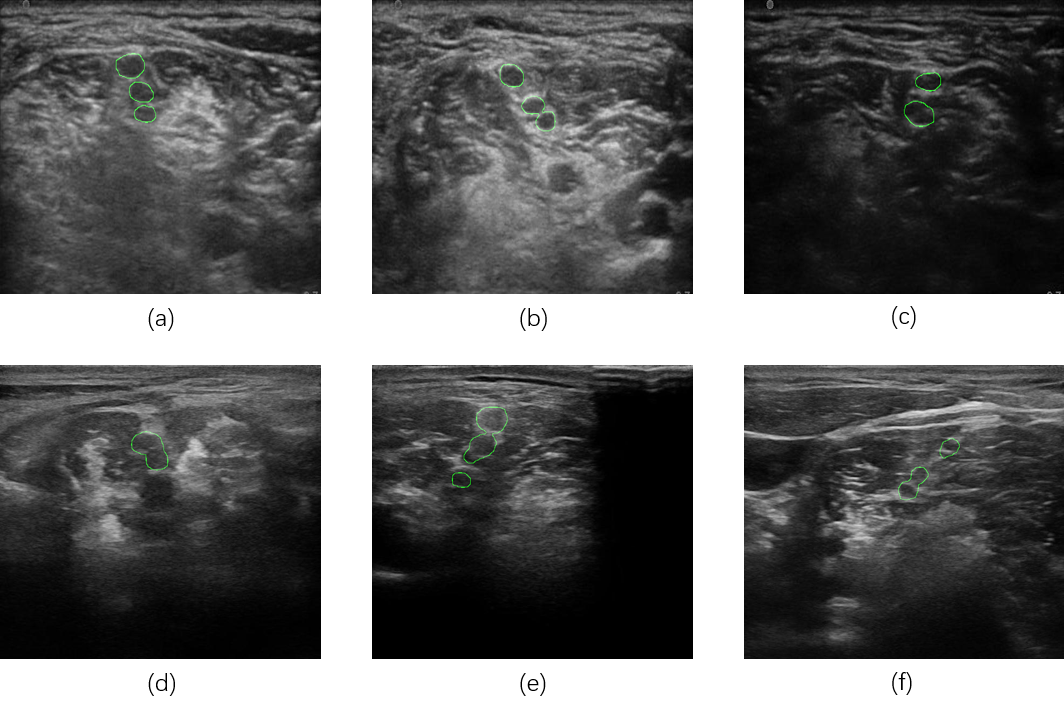}
		\caption{The final segmentation results of BPSegSys. (a), (b), and (c) are from YGY dataset images. (d), (e), and (f) are from BK3000 dataset images.}
		\label{result}
	\end{figure*}

	With this study, we built BPSegData to develop an AI BPSegSys. BPSegData is collected using two ultrasound devices, 185 BPUS (YGY dataset) collected using YGY equipment and 135 BPUS (BK3000 dataset) collected using BK3000 equipment, a total of 320 images. Three experienced doctors labeled the 320 images for BPSegData to ensure the accuracy of the brachial plexus trunks segmentation label. Using this dataset, we developed an AI system for the precise segmentation of the brachial plexus trunks, assisting the anesthesiologists to accurately identify the brachial plexus trunks for anesthesia. The developed BPSegSys can form an automated brachial plexus trunk segmentation pipeline, which consists of two modules: (1) BPUS preprocessing module, (2) brachial plexus trunks segmentation module. We cropped, enhanced, and augumented the original ultrasound images in the preprocessing module and used deep learning methods to segment the brachial plexus trunks in the segmentation module. In order to accurately evaluate the performance of BPSegSys and the reliability of the verification model, we used the 10-fold cross-validation method. Some final segmentation results of BPSegSys are shown in Figure \ref{result}.
	\subsection{Nerve segmentation results using BPSegSys}

	We separately analyze the YGY dataset, BK3000 dataset, and the mixed-dataset frome the two devices. YGY dataset has a total of 185 BPUS. In order to achieve 10-fold cross-validation, we randomly delete 5 images and used 180 images for cross-validation. We use random grouping to divide the YGY dataset into 10 folds. Four experiments are performed on each fold, using (1) original images, (2) modified loss function, (3) enchanted images,  and (4) combination of enchanted images and modified loss function (mixed-optimization). After completing the four experiments on each fold, we obtain the results shown in the Table \ref{BPSS_YGY}. The mixed-optimization method achieve the best results in the four experiments, and the intersection over union (IoU) reached 52.38\%. The second best result uses of enchanted-images, and IoU is 47.43$\%$. There is no significant difference between using the modified loss function and using the original images, IoU with 46.59$\%$ and 46.47$\%$ respectively. Considering the experimental results of each fold, we find that the mixed-optimization method achieve the best results on the average IoU and, at the same time, achieved the best results in each fold. Using the enchanted image method achieve the second best results in 5 groups. Therefore, we observe that the enhanced images contribute significantly to the BPSegSys performance.
	\begin{table*}[]
		\centering
		\caption{Performance of BPSegSys on YGY dataset in different experiments.}
		\resizebox{\textwidth}{!}{
		\begin{tabular}{@{}lllll@{}}
			\toprule
			Optimization method & Original images & Modified loss function & Enhanced images & Mixed-optimization \\ \midrule
			Fold 0 & 0.4281 & 0.3996 & 0.4020 & \textbf{0.4376} \\
			Fold 1 & 0.5905 & 0.5613 & 0.5480 & \textbf{0.6159} \\
			Fold 2 & 0.4204 & 0.4655 & 0.4340 & \textbf{0.5030}  \\
			Fold 3 & 0.4759 & 0.4941 & 0.4973 & \textbf{0.5488} \\
			Fold 4 & 0.4377 & 0.3989 & 0.4672 & \textbf{0.5216} \\
			Fold 5 & 0.4643 & 0.4741 & 0.4849 & \textbf{0.5130}  \\
			Fold 6 & 0.4695 & 0.4797 & 0.4742 & \textbf{0.5253} \\
			Fold 7 & 0.4732 & 0.4756 & 0.4582 & \textbf{0.5091} \\
			Fold 8 & 0.4754 & 0.4852 & 0.4943 & \textbf{0.5441} \\
			Fold 9 & 0.4124 & 0.4254 & 0.4830 & \textbf{0.5196} \\
			Average value       & 0.4647        & 0.4659               & 0.4743        & \textbf{0.5238}             \\ \bottomrule
		\end{tabular}
		}
		\label{BPSS_YGY}
	\end{table*}

	Similarly, we use the BK3000 dataset to conduct experiments and the results are shown in Table \ref{BPSS_BK300}. Similar to using the YGY dataset, BPSegSys using the BK3000 dataset also achieved the best results in the mixed-optimization method, reaching an IoU of 47.15$\%$, and achieved 6 best results in 10 folds. The modified loss function method achieved the second-best result, only 0.4$\%$ lower than the mixed-optimization method, reaching 46.75$\%$, and, out of 10 folds, 4 first places and 5 second places were obtained. Using the original image and enchanted-image methods, the IoU is 42.24$\%$ and 42.1$\%$, respectively. There is not much difference between the two methods, but compared with the previous two methods, the gap has reached more than 4$\%$. It can be seen that, on the BK3000 dataset, the modified loss function contributes noticeably to the BPSegSys performance.
	\begin{table*}[]
		\centering
		\caption{Performance of BPSegSys on BK3000 dataset in different experiments.}
		\resizebox{\textwidth}{!}{
		\begin{tabular}{@{}lllll@{}}
			\toprule
			Optimization method & Original images & Modified loss function & Enhanced images & Mixed-optimization \\ \midrule
			Fold0 & 0.4912 & \textbf{0.5541} & 0.5289 & 0.5533 \\
			Fold1 & 0.3376 & \textbf{0.5136} & 0.3464 & 0.4743 \\
			Fold2 & 0.4468 & 0.4621 & 0.4687 & \textbf{0.4688} \\
			Fold3 & 0.3068 & 0.3822 & 0.2978 & \textbf{0.3829} \\
			Fold4 & 0.4767 & 0.4868 & 0.4886 & \textbf{0.5206} \\
			Fold5 & 0.4453 & \textbf{0.4620} & 0.4026 & 0.4604 \\
			Fold6 & 0.5188 & \textbf{0.5391} & 0.5097 & 0.5169 \\
			Fold7 & 0.3465 & 0.3541 & 0.3504 & \textbf{0.3735} \\
			Fold8 & 0.4326 & 0.4706 & 0.3969 & \textbf{0.5017} \\
			Fold9 & 0.4213 & 0.4504 & 0.4203 & \textbf{0.4621} \\
			Average value       & 0.4224       & 0.4675                & 0.4210        & \textbf{0.4715}            \\ \bottomrule
		\end{tabular}
		}
		\label{BPSS_BK300}
	\end{table*}

	Subsequently, in order to test the generalization performance of BPSegSys, we examine BPSegSys on the mixed-dataset. In mixed-dataset, the ratio of the number of images between the YGY dataset and the BK3000 dataset is 37:27. We conducted four groups of 10-fold cross-validation experiments, and the mixed-optimization method still achieved the best results, with an IoU of 50.29$\%$. The result is between those using the BK3000 dataset and the YGY dataset alone, which is in line with expectations. The results of the other three experiments (original images, modified loss function, and enhanced images) are similar to the BK3000 dataset results and are shown in Table \ref{BPSS_mixed}. It can be seen that the modified loss function method has achieved significantly better results than the other methods.
	\begin{table*}[]
		\centering
		\caption{Performance of BPSegSys on mixed-dataset in different experiments.}
		\resizebox{\textwidth}{!}{
		\begin{tabular}{@{}lllll@{}}
			\toprule
			Optimization method & Original images & Modified loss function & Enhanced images & Mixed-optimization \\ \midrule
			Fold0 & 0.4772 & 0.5212 & 0.5016 & \textbf{0.5216} \\
			Fold1 & 0.4395 & 0.4762 & 0.4387 & \textbf{0.5136} \\
			Fold2 & 0.4056 & 0.4351 & 0.4079 & \textbf{0.4541} \\
			Fold3 & 0.4460 & 0.4674 & 0.4549 & \textbf{0.4891} \\
			Fold4 & 0.4209 & 0.4442 & 0.4021 & \textbf{0.4856} \\
			Fold5 & \textbf{0.4776} & 0.4603 & 0.4348 & 0.4743 \\
			Fold6 & 0.4306 & 0.4667 & 0.3945 & \textbf{0.4868} \\
			Fold7 & 0.5545 & 0.5544 & 0.5354 & \textbf{0.5588} \\
			Fold8 & 0.4830 & 0.5106 & 0.5102 & \textbf{0.5488} \\
			Fold9 & 0.4452 & 0.4961 & 0.4374 & \textbf{0.4966} \\
			Average value       & 0.4580        & 0.4832               & 0.4518        & \textbf{0.5029}            \\ \bottomrule
		\end{tabular}
		}
		\label{BPSS_mixed}
	\end{table*}
	
	\subsection{Doctors' manual segmentation results}
	As described above, we train BPSegSys and obtain the brachial plexus segmentation results of BPSegSys based on three dataset scenarios. Next, based on the manual segmentation results independently done by the three doctors (Doctor A, Doctor B, and Doctor C) on the brachial plexus, we analyze the doctors' performance in segmenting the brachial plexus and compare them with BPSegSys. We perform 10-fold cross-validation on the segmentation results of each doctor in the same way as before (YGY dataset, BK3000 dataset, and mixed-dataset). The results show that Doctor C achieves the best results in all three datasets, and the IoU is 52.05\%, 47.04\%, and 49.79\%, respectively. The results of the other two doctors (Doctor A, Doctor B) are relatively close. However, the best results among doctors are slightly worse than the experimental results of BPSegSys (52.38$\%$, 47.15$\%$, and 50.29$\%$). This shows that BPSegSys reaches a level comparable to that of senior doctors (Doctor A, Doctor B, and Doctor C) for brachial plexus identification. We also analyze the variance of three doctors' segmentation results on a given dataset (YGY dataset, BK3000 dataset, and mixed-dataset), which are 0.04623, 0.03939, and 0.04042, respectively. The variances of the experimental results of BPSegSys on the three datasets are 0.02806, 0.02764, and 0.02361, which are much smaller than the variances among the three doctors' results (Table \ref{doctor_result}). It can be seen from the variance results that the variance between the brachial plexus segmentation results of the three senior doctors are relatively large, indicating that the ultrasound diagnosis is highly subjective. In contrast, the variance between the BPSegSys segmentation results are relatively small, and the segmentation IoU performance results are slightly better than the best results among the doctors, demonstrating the stability and reliability of BPSegSys.

\begin{table*}[]
	\centering
	\caption{Doctor's segmentation results compared with BPSegSys.}
	\resizebox{\textwidth}{!}{
	\begin{tabular}{@{}lllllll@{}}
		\toprule
		& \multicolumn{4}{l}{IoU}                                          & \multicolumn{2}{l}{Variance}                 \\ \midrule
		& Doctor A & Doctor B & Doctor C                 & BPSegSys        & Doctor                    & BPSegSys         \\ \midrule
		YGY dataset   & 0.4288   & 0.4641   & \emph{\textbf{0.5205}} & \textbf{0.5238} & \textit{\textbf{0.04623}} & \textbf{0.02806} \\
		BK300 dataset & 0.4195   & 0.3929   & \emph{\textbf{0.4704}} & \textbf{0.4715} & \textit{\textbf{0.03939}} & \textbf{0.02764} \\
		Mixed-dataset & 0.4242   & 0.4323   & \emph{\textbf{0.4979}} & \textbf{0.5029} & \textit{\textbf{0.04042}} & \textbf{0.02361} \\ \bottomrule
	\end{tabular}
	}
	\label{doctor_result}
\end{table*}

\subsection{Doctors' segmentation results with the assistance of BPSegSys}
	After all the experimentation discussed above, we further study the role of BPSegSys in assisting doctors in brachial plexus identification. We selecte Doctor A's folds with lower scores in the YGY dataset and BK3000 dataset and then aske Doctor A to relabel the data considering the BPSegSys segmentation results. The time interval between the two labeling is more than three months, and the results are shown in Table \ref{contrast}. From Table \ref{contrast} we can see that, after referring to the BPSegSys segmentation results, the results of Doctor A on the two datasets have improved. Among them, the YGY dataset experiment achieves significant performance improvement, reaching 27\%; and the BK3000 dataset experiment has improved 3.9$\%$. Oversee, we observe that BPSegSys is helpful to anesthesiologists. Notice that BK3000 is an ultrasound device often used by Doctor A, and BPSegSys is able to assist the doctor in improving the segmentation performance, even though it's a small improvement (3.9\%). However, YGY is not a device often used by Doctor A and the doctor's segmentation results are significantly improved (27\%) with the aid of BPSegSys. In conclusion, BPSegSys is able to assist doctors in performing brachial plexus identification.

	\begin{table*}[]
		\centering
		\caption{Comparison of doctor's segmentation results before and after using BPSegSys.}
		\resizebox{\textwidth}{!}{
		\begin{tabular}{@{}lllll@{}}
			\toprule
			Dataset       & Fold  & Original result & Second result & Percentage of improvement \\ \midrule
			YGY dataset   & Fold 4 & 0.3728          & 0.4735        & 27\%                   \\
			BK3000 dataset & Fold 4 & 0.3973          & 0.4128        & 3.90\%                 \\ \bottomrule
		\end{tabular}
		}
		\label{contrast}
	\end{table*}
	
	\section{Discussion}
	
	Through this study, we have the following important observations. First, BPSegSys achieves or exceeds doctors' brachial plexus identification accuracy level in both the data collected by a single device and the mixed data (Table \ref{doctor_result}). It shows that BPSegSys has a good potential for clinical applications of ultrasound-guided nerve block anesthesia. Second, lack of consensus and subjectivity is a broad view of ultrasound diagnosis, that is, a lack of inter- and intra-reader repeatability~\cite{qian2021prospective}, which can be seen in the deviation in the identification of nerve location in nerve block (Table \ref{doctor_result}). BPSegSys can provide doctors with a relatively stable and accurate nerve identification result, and assist doctors in implementing accurate nerve identification for precise anesthesia. Third, we have shown that BPSegSys can improve doctors' identification performance of brachial plexus (Table \ref{contrast}), and BPSegSys can be used for training purposes for doctors.

\begin{figure*}[h]
		\centering
		\includegraphics[width = 11.5cm]{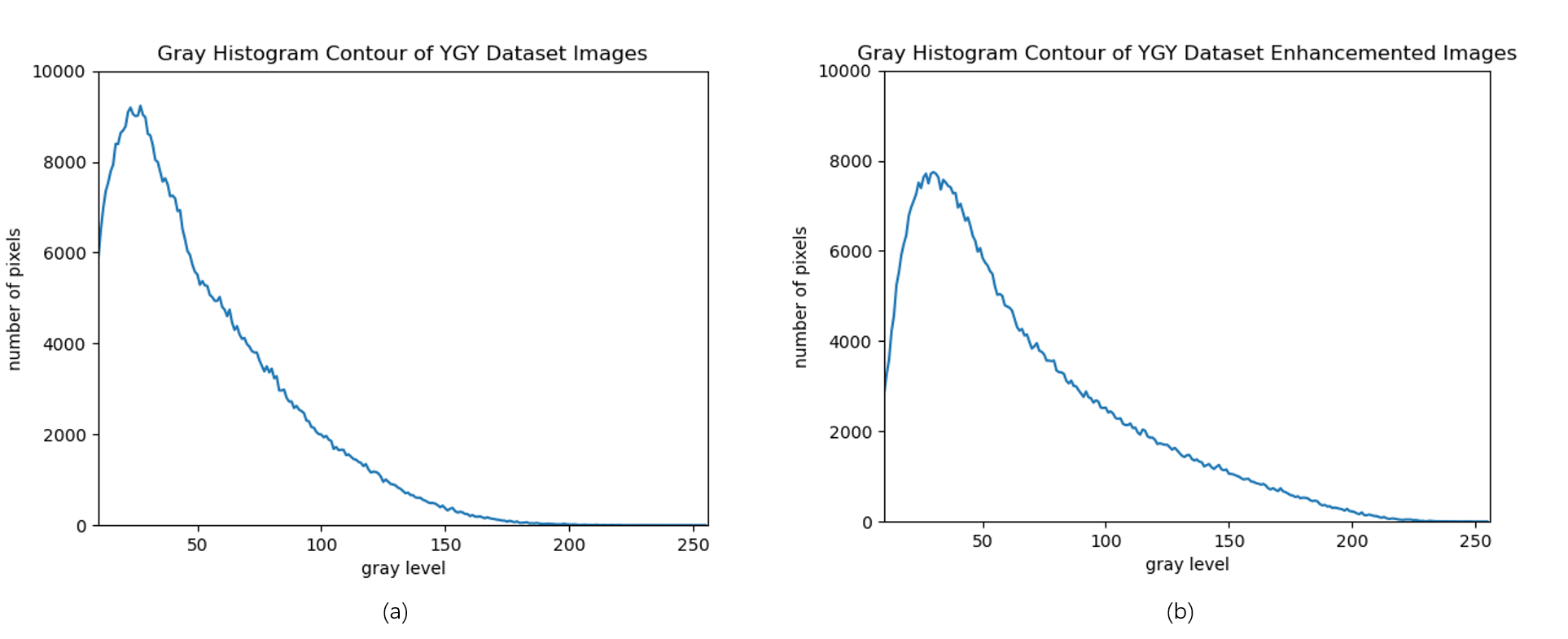}
		\caption{Histogram of YGY dataset image. (a) The original images' histogram, (b) The enhanced images' histogram.}
		\label{hist_YGY}
	\end{figure*}

\begin{figure*}[h]
		\centering
		\includegraphics[width=11.5cm]{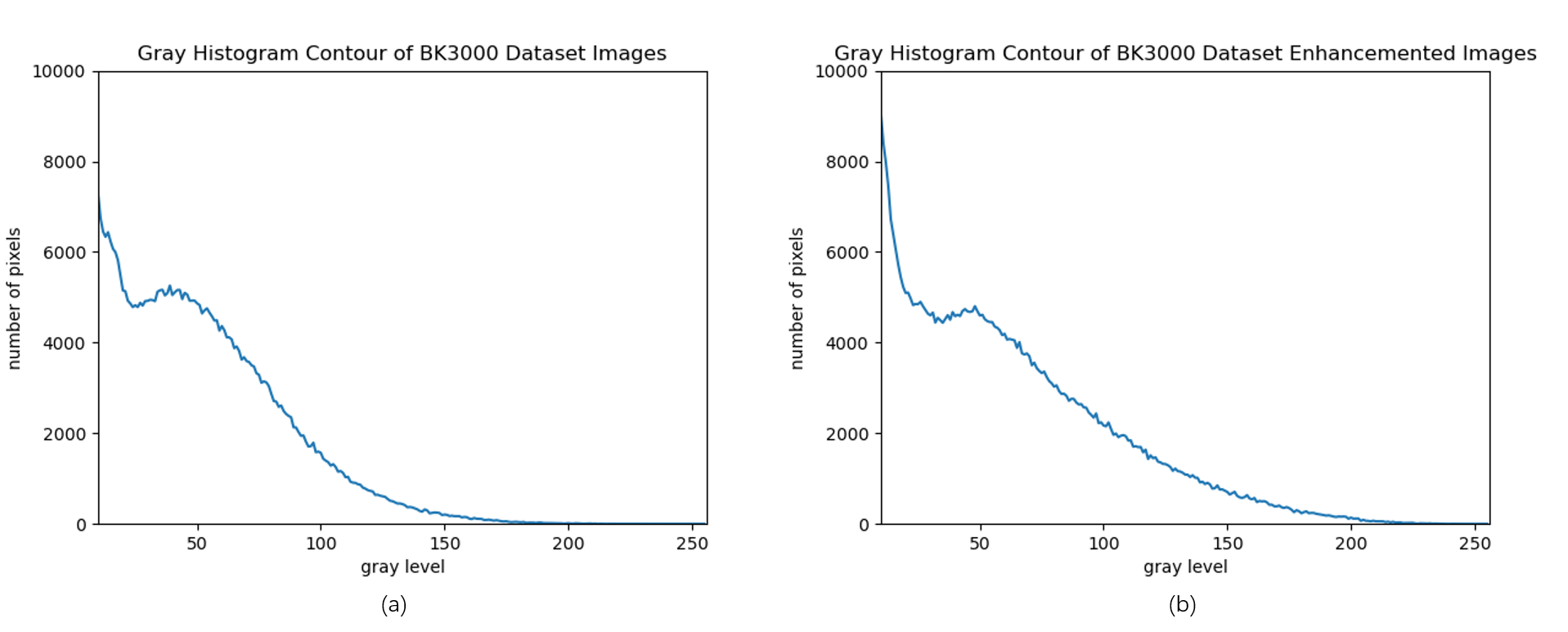}
		\caption{Histogram of BK3000 dataset image. (a) The original images' histogram, (b) The enhanced images' histogram.}
		\label{hist_BK3000}
	\end{figure*}

	We also investigate the influence of image features and network structure on BPSegSys. First, we observe the images of YGY dataset and BK3000 dataset and find that they have different grayscale features, as shown in Figure \ref{hist_YGY} and Figure \ref{hist_BK3000}. It can be seen that the histogram of the YGY dataset image is unimodal and distributed in areas with lower gray levels, and, in comparison, the histogram distribution of the BK3000 dataset image is relatively even. We perform image enhancement on YGY dataset and BK3000 dataset, respectively. The enhanced image of YGY dataset has a larger dynamic range of gray levels. Similarly, the gray value distribution of the enhanced image of BK3000 dataset is also improved, though not as significant in terms of the dynamic range and contrast. From the experimental results, we can see that on the YGY dataset, image enhancement can improve the performance of BPSegSys, but on the BK3000 dataset, it does not provide effective help. We also investigate the effect of the loss function on BPSegSys. BPSegSys with a modified loss function has better performance on all datasets. Notice that, after combining image enhancement and modified loss function (i.e., mixed-optimization), BPSegSys achieves the best results (Table \ref{doctor_result}). Therefore, our study demonstrates that improving the neural network structure and functions can improve the performance of BPSegSys in brachial plexus identification.
	
	There are still a few limitations in our work, which we hope to resolve in the future. First of all, although we observe that BPSegSys achieves a performance level comparable to that of doctors, other ultrasound devices are still needed to investigate the effectiveness of BPSegSys. Secondly, for elderly and obese patients, it is more difficult to observe the brachial plexus. In BPSegData, no separate statistics and experiments have been performed on elderly and obese patients, and we are unable to obtain the related performance results.
	
	In conclusion, we develope a brachial plexus ultrasound image segmentation system to assist doctors in identifying the brachial plexus trunks. The experimental data shows the system's feasibility and application potential in identifying the brachial plexus trunk. With the continuous advancement of deep learning technology, and continuous data collection, BPSegSys will be able to contribute more significantly for ultrasound-guided brachial plexus block.
	\section{Methods}
	
	\subsection{Datasets and label}
	Our research data comes from the Affiliated Hospital of Ningbo University and Ningbo No. 6 Hospital, collected by three experienced nerve block doctors from the location of the intermuscular sulcus, a total of 320 ultrasound images of the brachial plexus. The three doctors have been engaged in nerve block work for more than 10 years and have used ultrasound-guided nerve block for more than 7 years. The average annual number of patients for Doctor A is 700, and the average annual number of patients for Doctor B and for Doctor C is respectively 8,000. Among the 320 images, 135 images (BK3000 dataset) are collected by Doctor A using the high frequency linear 8870 probe on the BK3000 device, and the other 185 images are collected by Doctor B and Doctor C on the Sonosen S-Nerve device using Ultrassom HFL38xp probe acquisition (YGY dataset). After the data collection is completed, we label the data twice. (1) Each doctor individually uses Labelme to label the brachial plexus trunks and obtain the labeling results of each doctor based on their own knowledge and experience. (2) Three doctors label the brachial plexus trunks together, and the labeling results are used as the ground truth for BPSegData. Details of the doctors' experience, ultrasound devices, and operation are given in Table \ref{device}.
	
\begin{table*}[]
	\centering
	\caption{Details of the doctors' experience, ultrasound devices, and operation.}
	\resizebox{\textwidth}{!}{
\begin{tabular}{@{}llll@{}}
\toprule
                                  & Doctor A                   & Doctor B             & Doctor C             \\ \midrule
Ultrasound Device                 & BK BK3000                  & Sonosite S-Nerve     & Sonosite S-Nerve     \\
Ultrasound Probe                  & High Frequency Linear 8870 & HFL38xp              & HFL38xp              \\
Image Acquisition Region          & Intermuscular Groove       & Intermuscular Groove & Intermuscular Groove \\
Years of Working in Nerve Block   & 14 years                   & 10 years             & 11 years             \\
Years of Using UGNB               & 7 years                    & 8 years              & 8 years              \\
Annual Average Number of Patients & 700                        & 8000                 & 8000                 \\
Patients Age Range                & 18-80                      & 10-85                & 10-85                \\
\begin{tabular}[l]{@{}l@{}}Image Acquisition Process\\\\\\\\\\\\\\\\\end{tabular} &
  \begin{tabular}[l]{@{}l@{}}With the patient's head turned\\ to one side, the high-frequency\\ probe is used to translate upward\\ from the supraclavicular fossa.\\\\\\\\\\\end{tabular} &
  \begin{tabular}[c]{@{}l@{}}Turn the patient's head\\ toward the unaffected side,\\ use thehigh-frequency probe\\ to translate outward from\\ the plane of the cricoid\\ cartilage, and then move\\ up and down to confirm the\\ target.\end{tabular} &
  \begin{tabular}[c]{@{}l@{}}Turn the patient's head\\ toward the unaffected side,\\ use thehigh-frequency probe\\ to translate outward from\\ the plane of the cricoid\\ cartilage, and then move\\ up and down to confirm the\\ target.\end{tabular} \\ \bottomrule
\end{tabular}
}
\label{device}
\end{table*}
	\subsection{Image preprocessing}
	Before inputting an ultrasound image to BPSegSys, we performe some preprocessing on the image. Since the original image contains certain irrelevant information, such as the information frame of the ultrasound equipment software and the invalid area (i.e., the non-image area) of the ultrasound image. We first crop the image and extracte the area containing relevant information from the original image. For the image collected by the YGY device, we use (87, 47) as the upper left point and crop the image with a size of $510\times356$. The image collected by the BK3000 device contains two types of image interfaces. We use (278, 174) as the upper left point on the first interface to crop an image with a size of $553\times492$. The second interface’s upper left point is (165, 172), and the size of the captured image is $595\times529$. Finally, all cropped images are adjusted to $224\times224$ as the input of the deep learning network.
	
	We augmente our training data by applying horizontal flipping and random cropping. More specifically, we flip the image horizontally to double the amount of data. Then the above image is randomly cropped twice to obtain an image with a size of $224\times224$. In summary, we augmente the original data to 6 times, including the original images, 2 times cropped-images, fliped-images, and 2 times fliped-cropped-images.
	
	Due to the poor contrast of the ultrasound image, we performe a gray histogram analysis on the brachial plexus ultrasound image. As discussed in Section 3, we use contrast limited adaptive histgram equalization (CLAHE) method to improve image quality. We use the built-in toolkit of the image processing tool OpenCV-Python to perform CLAHE enhancement on ultrasound images, with clipLimit set to 1, and tileGridSize set to (8, 8).
	
	We use intersection-over-union (IoU) as the performance metrics in our experiments. The calculation of IoU is as follows,
	\begin{equation}
			\rm{IoU}=\dfrac{\text{Area of Overlap}}{\text{Area of Union}}
		\label{IOU}
	\end{equation}

	\subsection{Deep learning neural network development}
	At present, there are many high-performance image segmentation deep learning networks. Therefore, before we start this experiment, we first screen some deep learning segmentation networks, including U-Net~\cite{ronneberger2015u}, U-Net++~\cite{zhou2019unet++}, MedT~\cite{valanarasu2021medical}, Attention U-Net (Att U-Net)~\cite{oktay2018attention}. We chose the brachial plexus ultrasound image dataset~\cite{kaggle} with similar features as our dataset (BPSegData) as the test data. The dataset contains 5535 labeled ultrasound images, of which 2334 images contain brachial plexus nerves. We divide 2334 ultrasound images into the training set and test set according to 9:1 and conduct deep learning segmentation network training and testing. After testing, Att U-Net achieves the best results in all networks, and we finally selecte it as the network model used in this study. Att U-Net is an improved convolutional neural network based on U-Net. The attention gate is added to the upsampling of U-Net, which enhances the neural network's sensitivity to the region of interest.
	
	We use the image segmentation suite Paddleseg~\cite{paddleseg2019} tool developed based on the PaddlePaddle deep learning framework to implement ATT U-Net. We use NVIDIA RTX 2070 graphics processing unit for training and testing. The training set and test set are randomly divided in 9:1, and 10-fold cross-validation is used to evaluate the performance of the deep learning network. We use stochastic gradient descent (SGD) as an optimizer in a batch size of 4 with an initial learning rate of 0.01, which then decays every 1 iteration with 0.00001. 
	
	We first use original images and enhanced-images to train Att U-Net, respectively. Considering that the brachial plexus trunk occupies a relatively small proportion of the whole image, which belongs to the small target segmentation, we add the loss function Lovász hinge loss~\cite{berman2018lovasz} suitable for small target segmentation in Att U-Net. We combine Lovász hinge loss and cross-entropy loss to obtain a new loss function. New loss function = 0.02$\times$Lovász hinge loss + 1$\times$cross-entropy loss. Finally, we use the original images and enhanced-images to train the optimized Att U-Net, respectively. In training, we use the test set as the validation set of the model simultaneously and verify it every one epoch, retaining the model parameters with the highest IoU, that is, the optimal parameters during the entire training process.

\subsection*{Availability of data and materials}
Available upon request.
\subsection*{Authors' contributions}
Y.Y. initiated the project and oversaw all aspects of the project. Y.Y. and B.Z. conceived the project. Y.W. and D.T. designed the experiments. Y.W. conducted all experiments and analysed the experimental results. B.Z., L.K. and Jianlin Wang contributed to data collection, labelling and confirmation. D.T. assisted data labelling. B.G. and Jianhua Wang put forward helpful suggestions for the analysis of project. Y.W. and Y.Y. wrote the manuscript with input from all authors. All of the authors reviewed the manuscript.

{\scriptsize
	\bibliography{BrachialPlexusNerve}
	\bibliographystyle{ieeetr}}

\end{document}